\documentclass[12pt]{iopart}
\usepackage{iopams}  
\usepackage{graphicx}
\usepackage{xcolor}
\usepackage[colorlinks=true,linkcolor=magenta,hidelinks]{hyperref}
\usepackage{multirow}

\usepackage{xspace}
\makeatletter
\DeclareRobustCommand\onedot{\futurelet\@let@token\@onedot}
\newcommand{\@onedot}{\ifx\@let@token.\else.\null\fi\xspace}

\newcommand{\ie}{{i.e}\onedot\xspace}

\newcommand{\abinitio}{\textit{ab-initio}\xspace}
\def\fe2val{Fe$_2$VAl\xspace}

\newcommand{\dq}[1]{\lq\lq{}#1\rq\rq{}}%double quote
\newcommand{\rvec}{\mathbf{r}}

\newcommand{\kvec}{\mathbf{k}}

\let\baraccent=\= % rename builtin command \= to \baraccent
\renewcommand{\=}[1]{\stackrel{#1}{=}} % for putting numbers above =

\newcommand{\text}{\textrm}
\makeatother
\begin{document}

\title[Electronic structure of Zr-Ni-Sn systems: role of clustering and nanostructures...]{Electronic structure of Zr-Ni-Sn systems: role of clustering and nanostructures in Half-Heusler and Heusler limits}

\author{D.T. Do$^1$, S.D. Mahanti$^1$ and J.J. Pulikkoti$^2$}
%\homepage[ ]{http://www.msu.edu/~dodat}
%\address{Department of Physics and Astronomy, Michigan State University, East Lansing, MI 48824, USA}
\address{$^1$Department of Physics and Astronomy, Michigan State University, East Lansing, MI 48824, USA}
\address{$^2$CSIR - Network of Institutes for Solar Energy, CSIR - National Physical Laboratory, New Delhi 110012, India}
\ead{\mailto{dodat@msu.edu}, \mailto{mahanti@pa.msu.edu}}
\date{\today}

\begin{abstract}
Half-Heusler and Heusler compounds have been of great interest for several decades for thermoelectric, magnetic, half-metallic and many other interesting properties. Among these systems, Zr-Ni-Sn compounds are interesting thermoelectrics which can go from semiconducting half-Heusler (HH) limit, ZrNiSn, to metallic Heusler (FH) limit, ZrNi$_2$Sn. Recently Makogo et  al. [J. Am. Chem. Soc. 133, 18843 (2011)] found that dramatic improvement in the thermoelectric power factor of HH can be achieved by putting excess Ni into the system. This was attributed to an energy filtering mechanism due to the presence of FH nanostructures in the HH matrix. Using density functional  theory we have investigated clustering and nanostructure formation in ZrNi$_{1+x}$Sn ($0 \le x \le 1$) systems near the HH ($x=0$) and FH ($x=1$) ends and have found that excess Ni atoms in HH tend to stay close to each other and form nanoclusters. On the other hand, there is competing interaction between Ni-vacancies occupying different sites in FH which prevents them from forming vacancy nanoclusters. Effects of nano-inclusions on the electronic structure near HH and FH ends are discussed. 
%\keywords{Thermoelectric \and ab-initio \and electronic structure}
\end{abstract}

\maketitle

%\listoftodos\relax

\section{Introduction}
Half-Heusler and Heusler compounds have been of great interest for several decades for thermoelectric, magnetic, half-metallic and many other interesting properties. Development in the study of this family of compounds can be found in several review articles,%
\cite{galanakis_introduction_2006,galanakis_spin-polarization_2007,katsnelson_half-metallic_2008,picozzi_first-principles_2008,graf_heusler_2011,graf_simple_2011,casper_half-heusler_2012,bai_data_2013,galanakis_slater-pauling_2013} among which Graf \etal\cite{graf_simple_2011} gave a comprehensive overview of the field. Interesting examples of such compounds are M-Ni-Sn, M=Zr,Ti, and Hf, which are promising thermoelectric materials.%
\cite{aliev_metal-insulator_1987,aliev_narrow_1990,aliev_gap_1991,aliev_anomalous_1993,mestres_electron_1994,slebarski_electronic_1998,
hohl_efficient_1999,larson_structural_2000,shen_thermoelectric_2001,germond_thermoelectric_2010,kimura_vacancy_2010, qiu_effect_2010,kimura_effect_2011,lee_validity_2012,chen_effect_2013,romaka_peculiarities_2013,zou_electronic_2013} 
They can go from semiconducting half-Heusler (HH)\cite{slebarski_electronic_1998} limit, \ie ZrNiSn with cubic MgAgAs structure, to metallic Heusler limit (some times referred to as full-Heusler, FH), ZrNi$_2$Sn\cite{slebarski_electronic_1998} with cubic MCu$_2$Al-type structure. Because of their excellent thermoelectric properties, MNiSn and their solid solutions have attracted considerable experimental and theoretical interests over last several years.%
\cite{hohl_efficient_1999,%
shen_thermoelectric_2001,%
germond_thermoelectric_2010,%
chen_effect_2013,%
zou_electronic_2013,%
romaka_peculiarities_2013,%
sakurada_effect_2005,%
chaput_electronic_2006,%
kim_high_2007,%
wang_thermoelectric_2009,%
yu_high-performance_2009,%
hazama_improvement_2011,%
makongo_simultaneous_2011,%
makongo_thermal_2011,%
poon_half-heusler_2011,%
douglas_enhanced_2012,%
wang_chai_nanosized_2012,%
joshi_enhancement_2013,%
appel_effects_2013,%
birkel_improving_2013,%
kirievsky_phase_2013,%
yan_thermoelectric_2013,%
downie_enhanced_2013}

	Recently Makongo \etal\cite{makongo_simultaneous_2011, makongo_thermal_2011} (MK) have observed that large enhancements of both thermopower ($S$) and electrical conductivity ($\sigma$) and hence the powerfactor PF=$S^2\sigma$ of half-Heusler (HH) phase of (Zr,Hf)NiSn can be achieved at high temperatures through insertion of nanometer-scale FH coherent inclusions within the HH matrix, caused by adding a small concentration of extra Ni atoms. They refer these systems as HH($1-x$)/FH($x$) nanocomposites, which can be alternatively characterized as ZrNi$_{1+x}$Sn,  where $0<x<1$.
	% As shown in Fig.~\ref{fig.fhhh.peudo_pf}, 
In their work, Bi-doped HH(0.94)/FH(0.06) was found  to have a PF of 3.3~mW/mK$^2$ compared to 1.6~mW/mK$^2$ of Bi-Doped HH around 700~K.  Chai and Kimura\cite{wang_chai_nanosized_2012} (CK) also saw a large density of nanosized FH precipitates within HH matrix in slightly Ni-rich TiNiSn and found an increase in the power factor at high temperatures ($\sim$3.5~mW/mK$^2$ at 700 K compared to $\sim$2.5~mW/mK$^2$ at the same temperature for the parent compound TiNiSn\cite{kim_high_2007}). Both MK and CK suggested that an energy filtering mechanism proposed by Faleev and L\'eonard\cite{faleev_en_filter_2008} was operating in these systems due to the presence of metallic nano-structured inclusions in a semiconducting host. 
There is, however, no direct experimental evidence of such energy filtering caused by the FH nano-particles; more careful studies of electronic structure are needed to examine this idea.
The main idea behind the energy filtering mechanism proposed by Faleev and L\'eonard\cite{faleev_en_filter_2008} is the energy dependent scattering of the carriers by the metallic inclusion which filters out the low energy carriers from energy transport. Another possible way of incorporating energy filtering mechanism, particularly at high $T$, is to filter out either electrons or holes from energy transport, thereby, suppressing electron-hole cancellation.\cite{biswas_strained_2011} 

Romaka \etal\cite{romaka_peculiarities_2013} have carried out a thorough investigation of structural phase transitions in half-Heusler–-Heusler stannides (Zr,Hf)Ni$_{1+x}$Sn over the entire range of x=0$\rightarrow$1, and have shown how different physical properties change as one goes from HH to FH limit. Similar studies in  Ti-Co-Sn system was carried out by Stadnyk \etal\cite{stadnyk_isothermal_2005} who found that in TiCo$_{1+x}$Sn, electrical conductivity changed from metallic to semiconducting as they varied $x$ from 0 to 1. Romaka \etal\cite{romaka_peculiarities_2013} found that in ZrNi$_{1+x}$Sn, macroscopic phase separation of HH and FH phases took place in the region $0.3\leq x\leq 0.7$% (see Fig.~\ref{fig.fhhh.romaka_phase_dos}a)
. Their X-ray diffraction (XRD) measurements matched very well to the cubic MgAgAs structure for $x\leq 0.3$ and to the cubic MCu$_2$Al-type structure for $x\geq 0.7$. In the intermediate range $0.3<x<0.7$, a two-phase coexistence was seen. The presence or absence of nanostructures of one phase in the matrix of the other phase in the concentration ranges  ($x\leq 0.3$ or $x\geq0.7$) could not be established from their XRD measurements. One has to note that nanostructures of FH of 2--8~nm were observed in TEM measurements by Makongo \etal\cite{makongo_simultaneous_2011} for $x\sim0.06$. One of the interesting observations by Romaka \etal was the increase in the density of states at the Fermi energy (inferred from susceptibility measurement in the range $0<x<0.1$
, suggesting that new states appear at the Fermi-energy with increasing $x$. 
	
\begin{figure}
\includegraphics[width=\columnwidth]{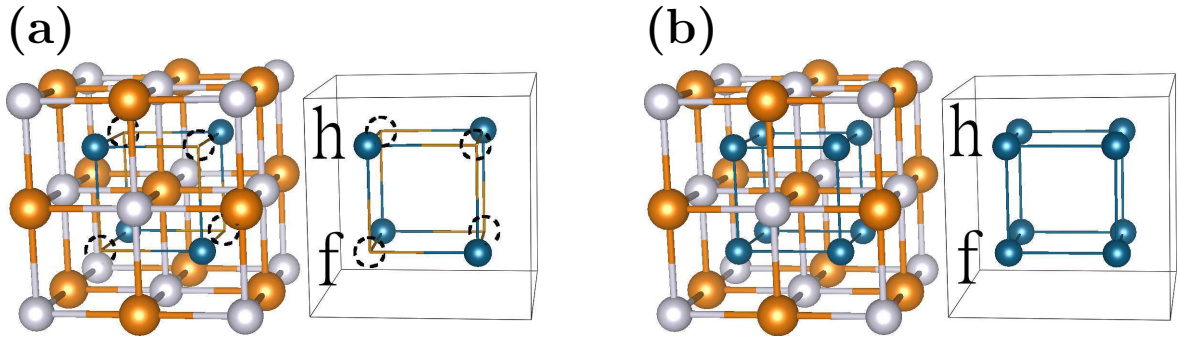}
\caption{\label{fig.hhfh.cryst}(Color online) Crystal structure of (a) Half-Heusler (HH) and (b) Full-Heusler (FH) compounds, where Zr (orange spheres) and Sn atoms (gray spheres) form a NaCl sub-lattice, inside which there are 8 small cubes.  HH is formed by filling every other cube (h-site) with Ni atoms (blue spheres), while FH is formed by filling all the cubes (both h- and f-sites). The empty-cube sites (quasi-particle of Ni-vacancy) in HH is presented by a dashed-line circle.}
\end{figure}

	One can visualize the structure of these HH and FH compounds as follows (Fig.~\ref{fig.hhfh.cryst}). One starts from a NaCl-type lattice of, say, ZrSn  consisting of two interpenetrating FCC sub-lattices. Inside the cubic unit cell of length $a$, there are 8 small cubes of length $a/2$. In ZrNiSn, the centers of 4 out of these 8 cubes are occupied by Ni atoms so that each Zr or Sn is tetrahedrally bonded to 4 Ni atoms. In ZrNi$_2$Sn, all the 8 small cube centers are occupied by Ni atoms (h-site). Thus one can tune from HH to FH or vice versa by varying the amount of Ni. The lack of mutual solubility of the HH and FH phases with similar crystal structures is quite remarkable in view of the fact that either adding Ni atoms to HH or removing Ni atoms from FH do not change the basic structure of the ZrSn matrix. %This has to do with local bondings, change in the electronic structure, and lattice relaxation as one goes from HH to FH limit.

	There are several fundamental questions that arise in the study of these mixed HH-FH systems. For example in ZrNi$_{1+x}$Sn, what is the nature of the phase diagram in the $T$ vs $x$ plot? What are the short and long range structural features and how they change with $x$? what are the electronic and lattice properties of the mixed system? There are some recent attempts to address these questions. For example, Romaka \etal\cite{romaka_peculiarities_2013} have looked at the electronic structure issue using Local Density Approximation (LDA)-single site Coherent Phase Approximation (CPA)\cite{akai_kkr_1989}, and Kirievsky \etal\cite{kirievsky_phase_2013} have studied the thermodynamics of Ni-rich TiNiSn compound, \ie the phase separation and energetics of anti-site defects using the supercell model. However, there are some limitations in these studies. For example, CPA used by Romaka \etal\cite{romaka_peculiarities_2013} cannot address the question of clustering and  nanostructures and the cubic supercell used by Kirievsky \etal\cite{kirievsky_phase_2013} in their electronic structure calculations is small ($1\times1\times1$ cubic unit cell). We will discuss some of their results and compare with ours in the appropriate limit. 
	
	In this work we address two of these basic questions. First, do the additional Ni atoms (or Ni vacancies) go randomly or arrange to form some sort of local ZrNi$_2$Sn (or ZrNiSn) nanoclusters? Second, what is the effect of these nanoclusters, if they form, on the electronic structure? In order to address these questions we have carried out electronic structure calculations using \abinitio density functional theory (DFT) and the supercell model\cite{defects} to describe the nanostructures. Our preliminary results were reported at "International Symposium on Clusters and Nanostructures - Richmond, VA, 2011".

This paper is organized as follows. In Sec.~\ref{sec.hhfh.method}, we discuss briefly details of electronic structure calculations and the model to describe the nanostructures. In Sec.~\ref{sec.hhfh.result}, we discuss the energetics of ZrNi$_{1+x}$Sn and their electronic structures, focusing on the nature of the gap states as one adds Ni atoms to ZrNiSn or the modification in the density of states near the Fermi energy as one adds Ni-vacancies to ZrNi$_2$Sn. Finally, in Sec.~\ref{sec.hhfh.sum}, we present a summary of our main findings and discuss possible further work in this area.
\section{\label{sec.hhfh.method}Methods of electronic structure calculation and modeling nanostructures}

It is well known that LDA/GGA calculations usually underestimate the band gaps in semiconductors\cite{defects,do_fe2val.2011,do_cusbse.2012}. However, the problem is opposite in the case of HH, MNiSn (M=Ti, Zr, Hf)\cite{ogut_bandgap_stability_1995,slebarski_electronic_1998,larson_structural_2000} where local density approximation (LDA) and generalized gradient approximation (GGA) calculations give band gaps about two times larger than that found from resistivity measurements\cite{aliev_narrow_1990} ($E_g=0.12, 0.18$ and $0.22$ for M=Ti, Zr and Hf respectively). The smaller band gap values have been ascribed to defects\cite{slebarski_electronic_1998,qiu_effect_2010}. We have tested the case of ZrNiSn with different improved approximations, namely modified-Becke-Johnson (mBJ)\cite{mbj} and hybrid functional (which includes non-local exchange partially) proposed by Heyd-Scuseria-Enzerhof (HSE06)\cite{hse06:1,hse06:2,hse06:3} methods. We have found that the band gap increased slightly from $\sim$0.48~eV in GGA to $\sim$0.51 eV with mBJ and to $\sim$0.58 eV with HSE06, indicating that GGA is sufficient to treat exchange-correlation in this system. In the FH limit, the systems are metallic and hence nonlocal effects are not expected to be important. Therefore, in this work total energy and electronic structure calculations were done using GGA. For the exchange-correlation potential we used the model suggested by  Perdew, Burke, and Ernzerhof (PBE).\cite{pbe} We employed projector-augmented wave (PAW) method\cite{bloch94, kresse99} as implemented in the VASP code.\cite{vasp1,vasp2,vasp3} with a plane-wave energy cutoff of 400 eV and an energy convergence criterion (between two successive self-consistent cycles) of $10^{-4}$ eV/unit cell.The charge densities and total energies were calculated self-consistently with an $8\times8\times8$ Monkhorst-Pack\cite{monkhorst76} $\kvec$-grid. The electronic density of states (DOS) was obtained using a denser $\kvec$-mesh of $12\times12\times12$.

In the present work, the lattice parameters of HH and FH are found to be 6.151~\AA\ and 6.321~\AA\ respectively, in good agreement with the experiments.\cite{slebarski_electronic_1998} The defective structures were obtained from these host structures and  ionic positions were then allowed to relax. To study how the electronic structure differs from the HH and FH limits as one adds and removes Ni atom(s) respectively, we have done calculations for ZrNi$_{1+x}$Sn using supercell method.\cite{defects} For convenience, we define $\delta$ as the defect concentration, where $\delta=x$ is the concentration of Ni atoms added to HH and $\delta=1-x$ is the concentration of Ni-vacancies in FH. The values of $\delta$ (or $x$) depends on the size of the supercell used in a calculation and the number of defects added. The supercell method has also been used by Kirievsky \etal\cite{kirievsky_phase_2013} for TiNi$_{1+x}$Sn, with a cubic $1\times1\times1$ cell (maximum 8 Ni atoms per unit cell) which gives large concentrations of defects ($\delta\geq 0.25$). Since, they focused primarily  on the thermodynamics of the HH-FH mixture and the quasi-binary-system phase diagram, total energy calculation using a small supercell is reasonable. However, because we are interested in nanostructure formation for small concentrations of extra Ni atoms or vacancies, we use a larger unit cell, mostly $2\times2\times2$ (containing 32 Zr, 32 Sn and up to 64 Ni atoms per supercell) with $\delta=m/32=0.03125 m,\, m=1,2,3,4$. For a few cases we have also used a $3\times3\times3$ supercell. We study the dilute limit of the defect by (i) starting from the HH structure and adding more Ni atoms and (ii) starting from the FH structure and removing some Ni atoms to create Ni-vacancies. We use the notation HH+$n$Ni (FH-$n$Ni) to denote a super cell of HH with $n$ additional Ni (FH with $n$ deficient Ni). Near the HH end, to understand how the FH nano-phase forms, all possible configurations of excess Ni-pairs were considered to find the lowest-energy (the most favorable) configuration. Once the preferred pair-configuration was found, the third Ni was added and we looked for the lowest energy configuration. An analogous procedure was applied for the FH end with Ni vacancies. 

In studying properties of defects in a material, it is useful to calculate their formation energies. There are extensive theoretical studies dedicated to this problem. Here we follow the work of Zhang \etal\cite{zhang_defect_cuinse_1998} and apply to HH-FH system where formation energy of a defect is given by
\begin{equation}
\Delta H_{f}(X_n)=\Delta E_{f}(X_n)+n\mu_{\text{Ni}},\label{eqn.form.enthanpy}
\end{equation}
where
\begin{equation}
\Delta E_{f}(X_n)=E(X_n)-E(0)+nE_{\text{Ni}}.\label{eqn.form.en}
\end{equation}
In these equations, $E(X_n)$ and $E(0)$ are the ground state energies of the $2\times2\times2$ supercell with and without defect $X_n$, $n$ is the number of Ni added ($n<0$) or removed ($n>0$) from the host-system (HH or FH), $E_{\text{Ni}}$ is the ground state energy per atom of Ni solid, and $\mu_{\text{Ni}}$ is the chemical potential of Ni. $\mu_{\text{Ni}}$ is usually considered as a parameter which varies from Ni-rich to Ni-poor environment. The limits of $\mu_{Ni}$ can be defined from thermodynamic equilibrium conditions which was carefully discussed by Zhang \etal\cite{zhang_defect_cuinse_1998}. In the present calculations, we will take $\mu_{Ni}=0$ and focus on $\Delta E_f(X_n)$. In an earlier calculation of defect formation energies by Kirievsky \etal\cite{kirievsky_phase_2013} a $1\times1\times1$ supercell was used, and the formation energies of the quasi-binary system,(TiNi$_2$Sn)$_{1-c}$(TiSn)$_c$, were calculated with respect to ground state energy of two end-systems. Furthermore, the physical meaning of their formation energies is unclear since TiSn is not a physical system.  Thus, a direct comparison of numbers between their results and ours is not meaningful. 
\section{\label{sec.hhfh.result}Results and Discussions}
\subsection{\label{subsec.hhfh.result.energetic} Formation energy and energetics of clustering}
\subsubsection{\label{subsubsec.hhfh.FHinHH}HH with excess Ni}
\begin{table*}
\caption{\label{tab.hhfh.energetics}Formation energies (eV) of HH-FH systems (for the lowest energy configurations)}
\def\arraystretch{1.5}
\begin{center}
\begin{tabular}{ccccccc}
\hline\hline
&HH+1Ni&HH+2Ni&HH+3Ni&HH+4Ni&FH-1Ni&FH-2Ni\\
\hline
$\Delta E_f(X)$&0.425&0.758&1.083&1.440&0.417&0.759\\
\hline\hline
\end{tabular}
\end{center}
\end{table*}

Table~\ref{tab.hhfh.energetics} gives the values of $\Delta E_f(X_n)$ needed to add or remove $n$ Ni atoms to the HH matrix or FH matrix respectively. All the formation energies are positive, \ie it costs energies to form HH-FH alloys at $T=0$K starting from each end. If we start from the HH end, it costs 0.423 eV to insert 1 Ni atom into one of the 32 available Ni sites. On the other hand, removing a Ni atom from FH (equivalent to create a vacancy at this end) is 0.417 eV. When more than one defect is created at either end, there is evidence of strong interaction between the defects. We will discuss this physics separately for the two ends, focusing more at the HH end due to its thermoelectric significance.

\begin{figure}
\includegraphics[width=\columnwidth]{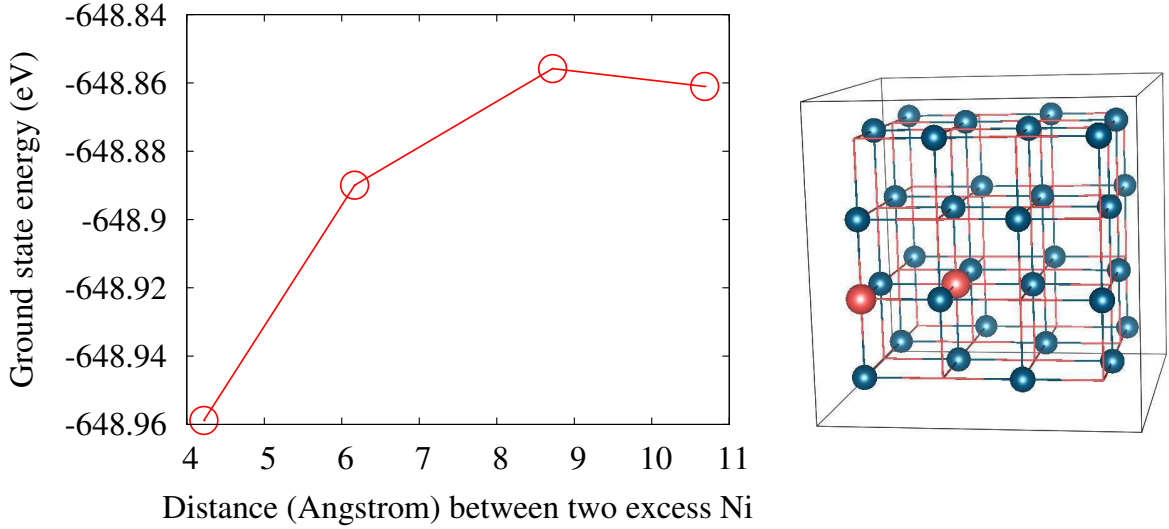}
\caption{\label{fig.fhhh.hh+2ni_en}(Color online) Left column presents energetics of two extra Ni in HH: energy vs. distance. Right column presents the preferable configuration of the excess Ni (orange sphere) (only Ni matrix (blue spheres) is shown).}
\end{figure}

With the first excess Ni (Ni$_1$) fixed, when the second Ni (Ni$_2$) is added to the HH matrix, there are several possible empty cube sites it can occupy. In Fig.~\ref{fig.fhhh.hh+2ni_en}(left column) we plot the ground state energy of HH+2Ni as a function of the distance between the Ni-pair (Ni$_1$ and Ni$_2$). It is clear that two Ni are attracted, since the energy is lowest at the smallest distance (4.22\AA). The preferred configuration of Ni-pair is shown in Fig.~\ref{fig.fhhh.hh+2ni_en}(right column). The ground state energy increases as the distance increases and then tends to saturate at large distances ($\sim$10\AA). The small decrease in the energy when the distance goes from $9\rightarrow10$\AA{} is due to the effect of the periodic supercell. Even with a $2\times2\times2$ unit cell, the size is only 12.3\AA$\times$12.3\AA$\times$12.3\AA, and there is an artificial periodicity of the defects. At large distances, the interaction of the excess Ni with their images in the neighboring unit cells counteract the intra-unitcell interaction. For a larger supercell, one should expect the energy to saturate at larger distances. The energy difference between the lowest- and the highest-energy configurations is $\sim$0.1~eV. From Table~\ref{tab.hhfh.energetics} we can estimate the binding energy of two Ni atoms as $\sim$0.088~eV (2$\times$0.423 eV $-$ 0.758 eV), close to 0.1~eV.

\begin{table*}
\caption{\label{tab.fhhh.hh+3ni_en}Ground state energy as a function of Ni$_3$'s relative position with respect to Ni$_1$-Ni$_2$, where $r=d_{\text{Ni}_3-\text{Ni}_1}$ and $\theta$ is the Ni$_3$-Ni$_1$-Ni$_2$ angle.}
\def\arraystretch{1.5}
\begin{center}
{\setlength{\tabcolsep}{0.4em}
\begin{tabular}{|c|cccc|c|cc|c|c|}
\hline\hline
$r$(\AA)&\multicolumn{4}{c|}{4.349}&6.151&\multicolumn{2}{c|}{7.533}&8.699&10.654\\\hline
$\theta$($^\circ$)&60&90&120&180&90&73.22&90&120&90\\\hline
E (eV)&-654.14&-654.10&-654.08&-654.06&-654.03&-654.03&-653.99&-653.99&-654.00\\
\hline\hline
\end{tabular}
}
\end{center}
\end{table*}

\begin{figure}
\includegraphics[width=\columnwidth]{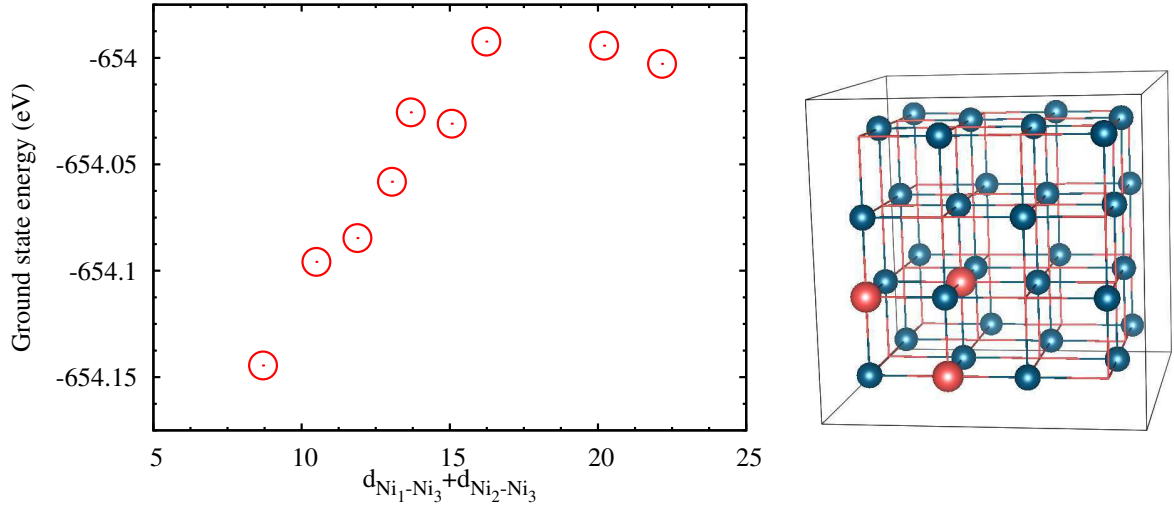}
\caption{\label{fig.fhhh.hh+3ni_en}(Color online) Left column presents energetics of HH+3Ni as a function of the sum of the distances of Ni$_3$ from Ni$_1$ and Ni$_2$ whose positions are predetermined by studying the energetics of a pair of excess Ni (FIG.~\ref{fig.fhhh.hh+2ni_en}). Right column shows the preferred configuration of the Ni$_3$ (only Ni matrix (blue sphere) is shown). The excess Ni are orange spheres, forming an equilateral triangle.}
\end{figure}

When the third Ni (Ni$_3$) is added to HH with two excess Ni, the problem is more complicated. As one can see from Table~\ref{tab.fhhh.hh+3ni_en}, the ground state energy depends on both the distance from Ni$_3$ to Ni$_1$ ($r$) and the Ni$_3$-Ni$_1$-Ni$_2$ angle ($\theta$). However, if one plots the ground state energies as a function of the sum of the distance of Ni$_3$ from Ni$_1$ and Ni$_2$ (as shown in Fig.~\ref{fig.fhhh.hh+3ni_en}(left column)), one gets a much simpler picture in which energy increases with increasing distance and saturates at large distance (similar to the case of Ni-pair discussed above). The difference between the highest- and lowest-energy configurations is about 0.14~eV. The lowest energy of HH+3Ni is found where Ni$_3$ is closest to both Ni$_1$ and Ni$_2$ so that the three Ni atoms occupy three corners of a tetrahedron, forming an equilateral triangle, as shown in Fig.~\ref{fig.fhhh.hh+3ni_en}(right column). Three excess Ni atoms tend to come close to each other and form a nano-cluster (a trimer) in the HH matrix. 

\begin{figure}
\includegraphics[width=\columnwidth]{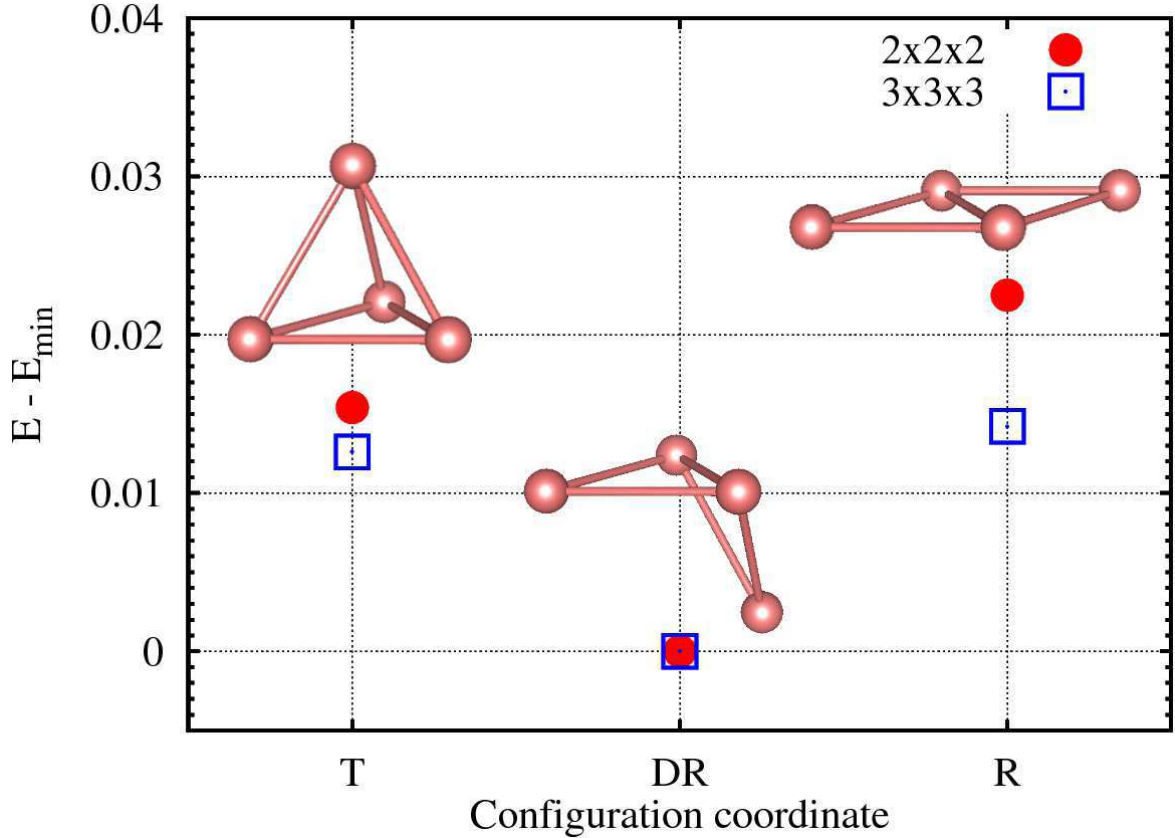}
\caption{\label{fig.hh+4ni_en}(Color online) Total energies of selected configuration of (Ni)$_4$ clusters in HH matrix (with respect to the energy of the most stable configuration) calculated for $2\times2\times2$ and $3\times3\times3$ supercell}
\end{figure}

Now let us look at the case of four excess Ni (note that a smallest unit of FH needs four excess Ni atoms). When the 4th Ni is added to the (Ni)$_3$ trimer (equilateral triangle geometry), the question is where does the 4th Ni atom go? This question has been extensively studied in the case of (Ni)$_4$ atomic clusters in free space.\cite{menon_tb_md.1994,Reuse_Nicluster.1995,Jena_energetics.1997,valeri_Nien.2004,futscheck_abinitio.2006,lu.2007}. Study also shows that Ni cluster's structure is easily deformed by contamination.\cite{Deshpande.2005,petkov.2006,deshpande.2007,chikaoui.2011}  Clearly the constrained environment of ZrNiSn matrix will play an important role in the structure of (Ni)$_4$ clusters.  For  (Ni)$_4$ in free space there are three competing geometries; a tetrahedran (T), a rhombus (R), a deformed rhombus (DR) which is in between T and R. (see Fig.~\ref{fig.hh+4ni_en}). It is believed that for (Ni)$_4$ clusters in free space, the T is the lowest energy structure. If the T structure in ZrNiSn matrix has the lowest energy then it would form an embryonic FH cell. However we find that neither T nor R have the lowest energy, rather the intermediate DR structure has the lowest energy (Fig.~\ref{fig.hh+4ni_en}). In this structure all of the Ni-Ni bonds except one have equal (short) lengths with one long bond. Note that in the T structure all Ni-Ni bonds are same (short bonds).

For the $2\times2\times2$ supercell, the energy difference between T and DR structures E(T)-E(DR)=0.015~eV and the difference between R and DR, E(R)-E(DR)=0.022~eV. One may suspect that 4 excess Ni atoms in $2\times2\times2$ supercell are too many and the effect of the periodic boundary condition is strong. In order to circumvent this situation we increase the supercell size to $3\times3\times3$ (allowing more local relaxation and reducing the effect of periodic boundary condition), we find that the energy different between T, R and DR decrease, especially between R and DR, indicating that planar structures relax more than the compact tetrahedral structures. 

The above results show that excess Ni in HH tend to stay close to each other, forming clusters. However, upto four excess Ni atoms, the formation of nanostructure of FH is not favorable in both $2\times2\times2$ and $3\times3\times3$ supercell. These clusters ($n=1-4$), however, may form nuclei for larger clusters which include the FH structure as observed in experiments. It will be interesting to study the evolution of the cluster of excess Ni atoms in HH for clustering size larger than four.

\subsubsection{\label{subsubsec.hhfh.HHinFH}FH with deficient Ni}
At the FH end, one should note that there are two different Ni-sites (Fig.~\ref{fig.hhfh.cryst}): \dq{h}-site (h) at (1/4,1/4,1/4) and its equivalents, which are occupied by Ni in HH, and \dq{f}-site (f) at (3/4,3/4,3/4) and its equivalents, which are empty in HH. When removing Ni, the Ni-vacancy can be located at either of these two types of sites. 

\begin{figure}
\includegraphics[width=\columnwidth]{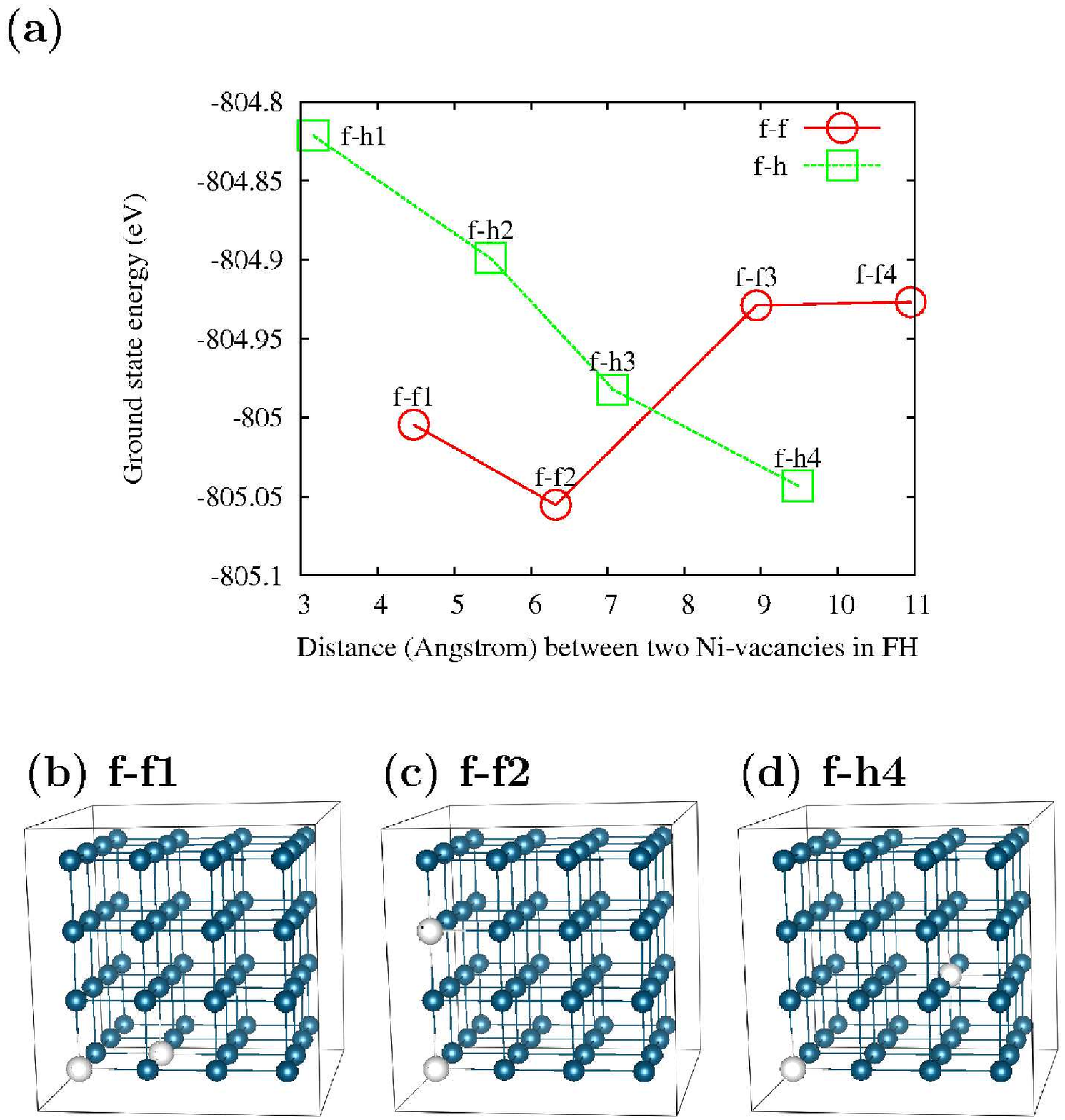}
\caption{\label{fig.fhhh.fh-2ni_en}(Color online) The top figure shows energetics of FH with two Ni-vacant sites (white spheres); the bottom shows structures of the three most preferable configurations, which have ground state energies close to each other: (a) f-f1, (b) f-f2, and (c) f-h4. Only Ni matrix (blue spheres) is shown.}
\end{figure}

When one Ni is removed, due to symmetry, f- and h-sites are equivalent. The energy cost for removing 1 Ni out of a $2\times2\times2$ unit cell is $\sim0.417$ eV. Similar to the case of HH+$n$Ni, the energy cost to remove the second Ni (to add the second vacancy) is smaller (by 0.045 eV) than that needed to remove the first Ni (Table~\ref{tab.hhfh.energetics}). We chose the first vacancy (V1) to be at a f site and the second one can be at either h or f site. We denote the pair of vacancies by the name of the sites which they occupy, \ie  f-f and f-h. Note that f-f and h-h are equivalent. In Fig.~\ref{fig.fhhh.fh-2ni_en}(a), we plot the energy of the supercell with a pair of vacancies as a function of the distance between them, considering f-f and f-h separately. In Fig. \ref{fig.fhhh.fh-2ni_en}(b,c,d) we show three possible di-vacancy configurations. Fig.~\ref{fig.fhhh.fh-2ni_en}(a) shows that f-f and f-h are  energetically different. 

For the f-h case, the interaction between two vacancies is repulsive,\ie the total energy is the lowest for the largest distance. The relation between energy and distance is almost linear. The difference between the lowest and the highest energies is about $\sim 0.22$~eV. 
For the f-f case, in contrast, the interaction between the two vacancies is, in general, attractive, the smaller is the distance, the lower is the energy. There is, however, a competing effect between repulsion and attraction at small distance, which makes the energy higher at the smallest distance, 4.47\AA\ (f-f1), than at the larger one, 6.32\AA\ (f-f2) (-805.00~eV for f-f1 compared to -805.055~eV for f-f2, as shown in Fig.~\ref{fig.fhhh.fh-2ni_en}(a)). The energies saturate at large f-f distance, however the actual value of the saturation energy is affected by the periodicity of the supercell. The structure of f-f2 (Fig.~\ref{fig.fhhh.fh-2ni_en}(c)) has 2 Ni-vacancies distributed evenly at every other site along the cubic axis (say $x$), forming a quasi one dimensional chain. 

The structure consisting of quasi 1-dimensional chains of vacancies discussed above is similar to the configuration (i) of Kirievsky \etal\cite{kirievsky_phase_2013}, except that they put one vacancy in a $1\times1\times1$ simple cubic super cell; thus the vacancies form a 3-dimensional cubic lattice. In their work, Kirievsky \etal\cite{kirievsky_phase_2013} calculated electronic structures of 11 configurations with different numbers of Ni occupying eight cubic sites available for Ni atoms in a TiSn matrix and studied their energetics using a quasi binary model (TiNi$_2$Sn)$_{c}$(TiSn)$_{1-c}$. They defined the formation energy as
\begin{equation}
\Delta U=E^{mixture}_{tot}-(c\cdot E^{TiNi_2Sn}_{tot}+(1-c)\cdot E^{TiSn}_{tot}),\label{eqn.kirievsky_energy}
\end{equation}
where $E_{tot}$ is the total energy per cubic cell of corresponding systems. They found that the configuration (i), $c=0.875$, had negative free-energy of formation whereas most of the other configurations (except $c=0.5$ which corresponds to Half-Heusler phase TiNiSn) have positive free-energy of formation. As mentioned earlier, one should note that the definition of formation energy used by Kirievsky \etal\cite{kirievsky_phase_2013} is different from the definition used in the present work. Furthermore, they have used a $1\times1\times1$ supercell where the periodic cell constraints are more important than ours.\footnote{However, it is reasonable to use a 1$\times$1$\times$1 supercell here because Kirievsky \etal\cite{kirievsky_phase_2013} used DFT energies to get effective parameters for their finite $T$ calculations.} Thus, it is not possible to directly compare the  values of formation energies between these two works. 

It is interesting that the energy difference between f-f2 (the lowest energy among f-f) and f-h4 (the lowest among f-h) is about $\sim0.012$~eV which is small (see Fig.~\ref{fig.fhhh.fh-2ni_en}). Thus, f-f2 and f-h4 can coexist at $x=0.9375$. This suggests a competition between different structure associated with excess Ni-vacancies in the FH matrix, meaning HH nanostructures are unlikely to form at the FH end for low concentration of Ni-vacancies.

\subsection{\label{subsec.hhfh.result.electronic}Electronic structure}
\begin{figure}
\begin{center}
\includegraphics[width=.9\columnwidth]{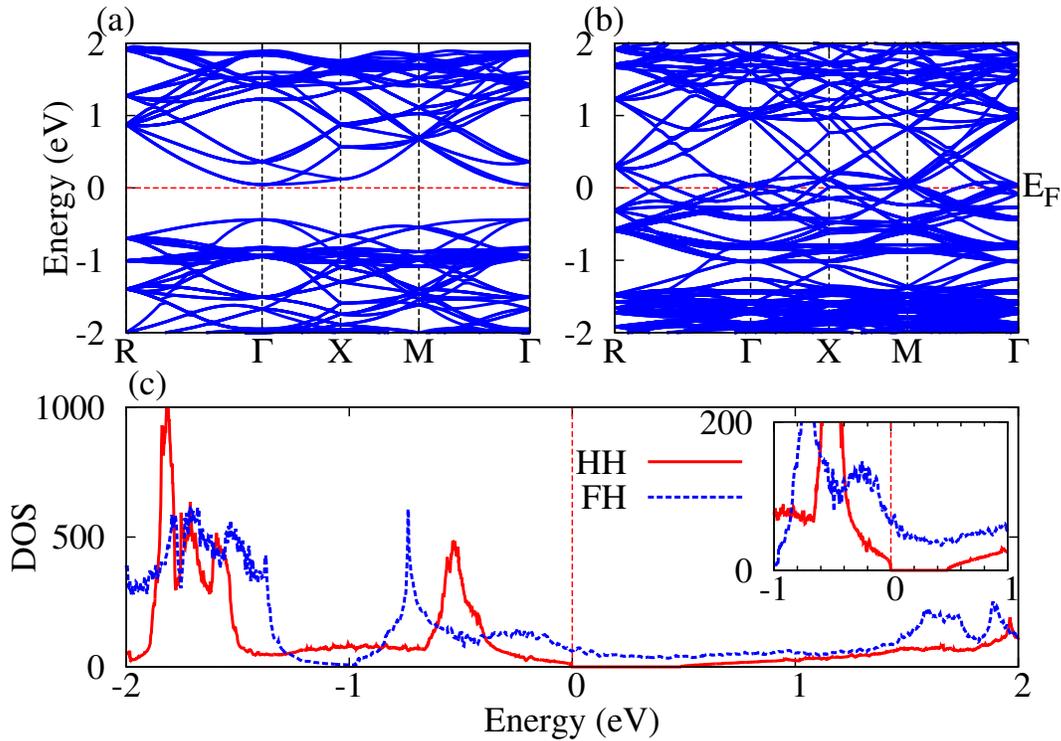}
\end{center}
\caption{\label{fig.fhhh.pureband}Band structures of (a) HH and (b) FH in the Brillouin zone (BZ) of the $2\times2\times2$ cubic supercell; and (c) the corresponding density of states (DOS) within energy range [-2,2] eV; the inset show the DOS in the range [-1,1] eV, emphasizing the fundamental difference between HH and FH near the Fermi level: the former is a metal whereas the latter is a semiconductor, with a band gap $\sim0.5$ eV.}
\end{figure}

Before we discuss the electronic structure of HH-FH mixture, let us first review the electronic structures of the two end compounds ZrNiSn and ZrNi$_2$Sn. Electronic structures of ZrNiSn and ZrNi$_2$Sn are well understood and were extensively studied both theoretically\cite{ogut_bandgap_stability_1995,slebarski_electronic_1998} and experimentally\cite{aliev_narrow_1990,aliev_gap_1991,ogut_bandgap_stability_1995}. ZrNiSn is a semiconductor with an indirect band gap of $\sim0.5$~eV\cite{ogut_bandgap_stability_1995,slebarski_electronic_1998,larson_electronic_1999} between  $\Gamma$ and $X$ points in the FCC Brillouin Zone (BZ) (not shown). The band gap is formed mainly from the hybridization of Zr-$d$ and Sn-$p$ orbitals with a mixing of some Ni-$d$ orbitals. The band gap formation was carefully discussed by \"O\ifmmode \breve{g}\else \u{g}\fi{}\"ut and Rabe\cite{ogut_bandgap_stability_1995} and Larson \etal\cite{larson_electronic_1999}. The top three valence bands are degenerate at the $\Gamma$ point but split as one goes away from it. One of these three bands has light effective mass while the other two have heavier mass. On the other hand, the lowest conduction band is non-degenerate with anisotropic effective mass. All Ni-$d$ and Sn-$s$ bands are fully occupied, the latter is located at about 15~eV below the valence band maximum. ZrNi$_2$Sn, on the other hand, is a paramagnetic metal with finite density of states at the Fermi level. In this system, the $d$-levels of Ni strongly hybridize with the Sn-$p$ and Zr-$d$ levels. The states below the Fermi energy (0 eV) in the energy range (-3 to 0 eV) are dominated by the Ni-$d$ orbitals with significant mixing with Zr-$d$ orbitals near the Fermi energy. In fact, such a mixing gives rise to peaks in the DOS at $\sim0.5$ eV, caused by Van-Hove singularity, which plays an important role in the high-$T_c$ superconductivity of the related compound ZrNi$_2$Ga.\cite{winterlik.superonductor.2008}

Fig.~\ref{fig.fhhh.pureband} gives the band structures of ZrNiSn (Fig.~\ref{fig.fhhh.pureband}a) and ZrNi$_2$Sn (Fig.~\ref{fig.fhhh.pureband}b) in the Brillouin zones (BZ) of $2\times2\times2$ simple-cubic supercell, and the corresponding density of states (DOS) per supercell (Fig.~\ref{fig.fhhh.pureband}(c)). Going from a fcc primitive unit cell to $2\times2\times2$ simple-cubic supercell, due to band folding, the band structure of ZrNiSn shows a direct (instead of indirect as in FCC (not shown) band gap of $\sim0.5$ at the $\Gamma$ point, in good agreement with earlier works\cite{ogut_bandgap_stability_1995,slebarski_electronic_1998,larson_electronic_1999}.  Fig.~\ref{fig.fhhh.pureband}(c) shows comparison between the DOS of the two compounds, near the Fermi level ($E_F=0$ eV).  The inset shows the DOS in the vicinity of Fermi level. Our calculation finds DOS($E_F$)=2.1 states/eV per formula unit, corresponding to electronic specific heat coefficient $\gamma$=5 mJ/mol K$^2$ which is about two times smaller the experimental value\cite{boff_specificheat.1996}. This discrepancy between theoretical and experimental estimations of $\gamma$ is well-known for Heusler compounds and may be ascribed to the renormalization by electron-phonon or electron-magnon (for magnetic systems) interactions.\cite{boff_specificheat.1996}

\begin{figure}
\includegraphics[width=\columnwidth]{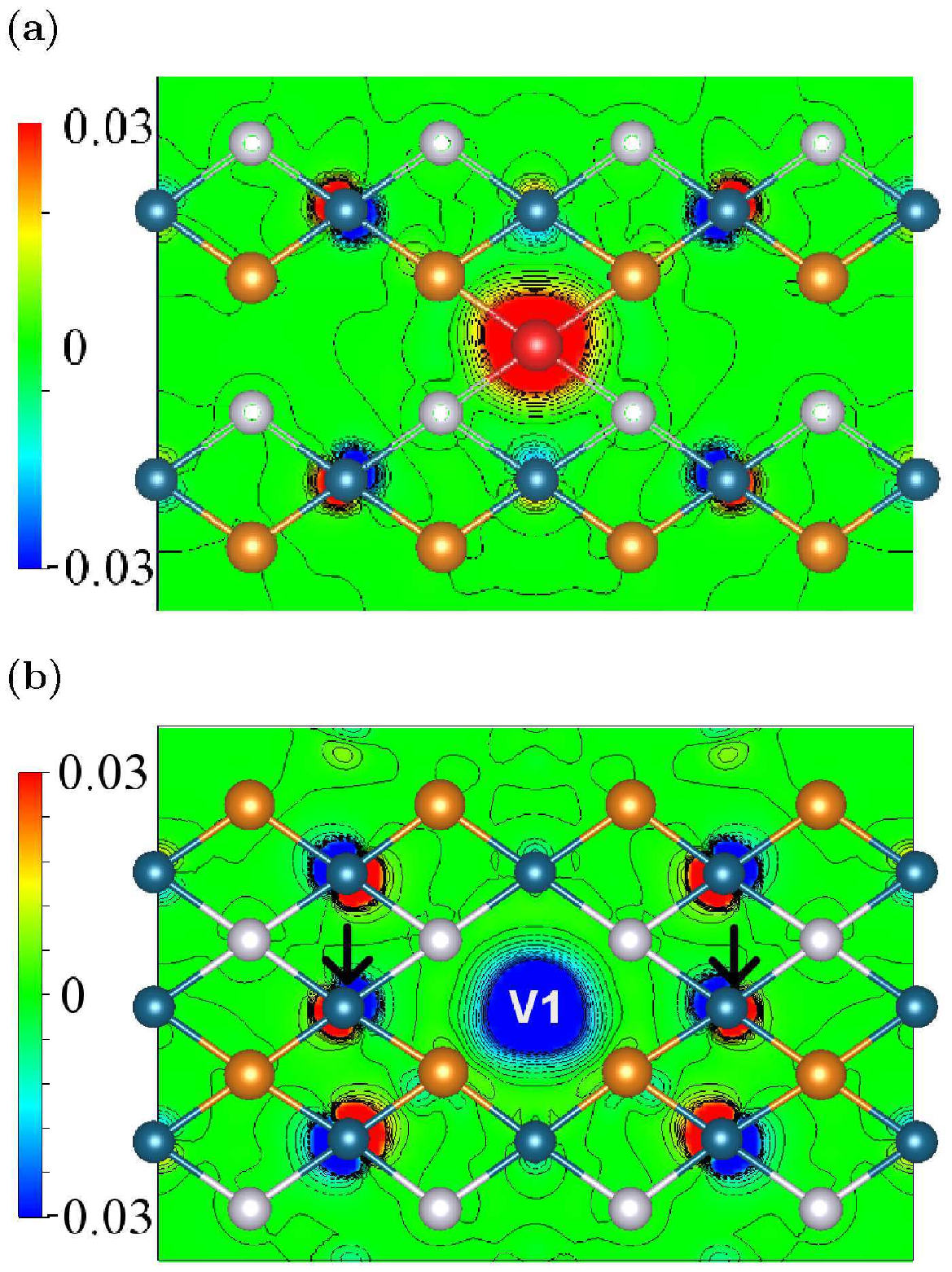}
\caption{\label{fig.hhfh.1ni.chargediff} Charge difference between (a) HH+1Ni and (b) FH-1Ni and their parent compounds (HH and FH), where blue color indicates charge depletion and red color indicates charge accumulation. The figure show the contour plot in the (-110) plane. Charge difference is defined as $\rho^X(\rvec)-\rho^0(\rvec)$, where $\rho^X(\rvec)$, and $\rho^0(\rvec)$ are charge densities of defective and pure systems respectively. Host-Ni atoms are blue spheres, excess Ni is red sphere, Zr are brown spheres, Sn are gray spheres.}
\end{figure}

When one Ni (Ni$_1$) is introduced into the HH matrix (or removed (V1) from FH), there is a change in the electronic structure and a redistribution of the charge density with respect to the host system (HH or FH). To analyze this charge redistribution, we calculate the charge density difference (charge-difference) $\Delta\rho=\rho^X(\rvec)-\rho^0(\rvec)$, where $\rho^X(\rvec)$ and $\rho^0(\rvec)$ are the total charge densities of the defective system and the corresponding host system. Fig.~\ref{fig.hhfh.1ni.chargediff} shows the charge difference contour on the ($\bar{1}$10) plane for HH+1Ni (Fig.~(\ref{fig.hhfh.1ni.chargediff}a) and FH-1Ni (Fig.~(\ref{fig.hhfh.1ni.chargediff}b), where blue indicates charge depletion and red indicates charge accumulation. 

For HH+1Ni, the additional Ni introduces 10 more valence electrons to the HH system, these electrons are localized around Ni$_1$ (Fig.~\ref{fig.hhfh.1ni.chargediff}a). The electrons at the excess Ni site cause rearrangement of charge in the neighboring region. However, this rearrangement occurs mostly at the Ni-sites. Electrons tend to move further away from Ni$_1$. In Fig.~\ref{fig.hhfh.1ni.chargediff}a one can clearly see that the blue lobes (representing charge depletion) at the Ni-sites are directed towards Ni$_1$, whereas the red lobes (charge accumulation) at the same sites are directed away from Ni$_1$. In Fig.~\ref{fig.hhfh.1ni.chargediff}(a) one sees that the Ni atoms of the host matrix (HH) are connected to Ni$_1$ indirectly through Zr or Sn. Let us call these connections Zr- and Sn-channels. With these channels defined, one can realize that the charge rearrangements are not the same at all host-Ni, they are larger on Ni which interacts with the Ni$_1$ through only one channel, either Sn-channel or Zr-channel, where Ni-Zr(Sn)-Ni$_1$ form a straight line. The charge rearrangements on other Ni-sites, which connect to Ni$_1$ through two channels of  the same type (either two Sn-channels or two Zr-channels), are smaller even though they may be at shorter distances from Ni$_1$. This clearly indicates the anisotropic nature of bonding between two Ni atoms in the system.

A similar picture also emerges in FH-1Ni (Fig.~\ref{fig.hhfh.1ni.chargediff}b), except that the removal of Ni (V1) takes away electrons. A subtle difference between FH-1Ni and HH+1Ni is that there is an interesting charge rearrangement on Ni indicated by arrows in Fig.~\ref{fig.hhfh.1ni.chargediff}b (hereafter it is called Ni$_x$). Ni$_x$
indirectly connects to V1 through two channels, one Zr-channel and one Sn-channel, the Ni$_x$-Sn(Zr)-V1 angle is about $\sim110^\circ$. The charge rearrangement on Ni$_x$ site does not occur along the Ni-V1 direction but along the Ni-Sn direction, as shown in Fig.~\ref{fig.hhfh.1ni.chargediff}b. The blue lobe around Ni$_x$ is closer to V1 than the red lobe, opposite to those around the other Ni atoms with linear link Ni-Sn(Zr)-V1, whose red lobes are closer to V1. This difference in charge rearrangement is most likely the reason for the difference in energetics of FH-$n$Ni from HH+$n$Ni and is a reflection of different local bondings in FH and HH systems. 

\begin{figure}
\centering
\includegraphics[width=.9\columnwidth]{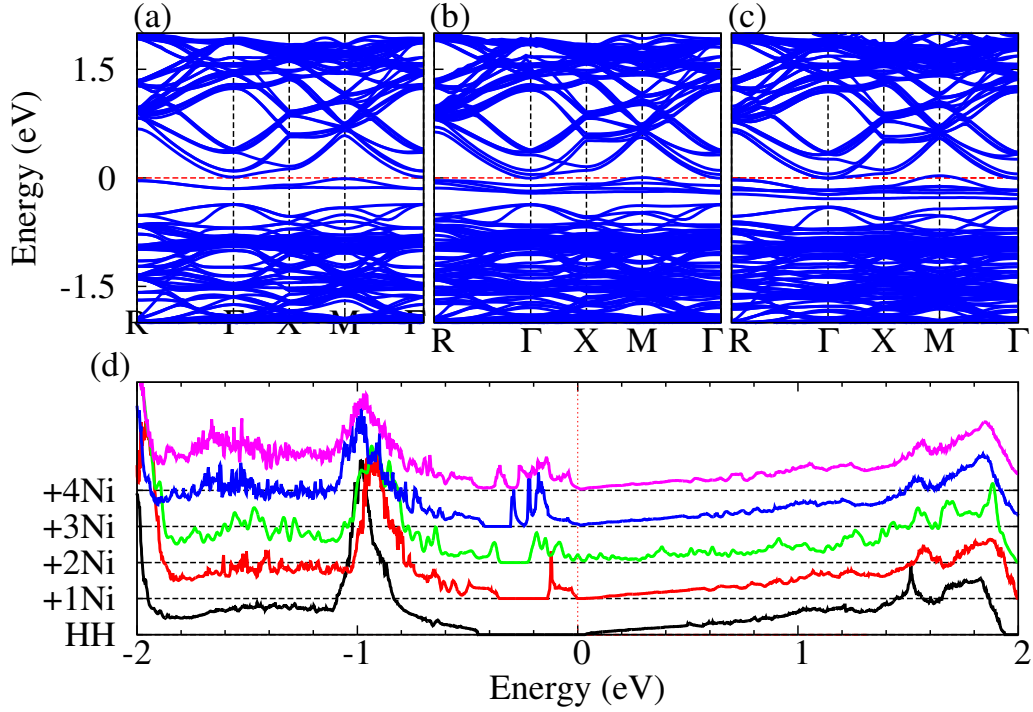}
\caption{\label{fig.fhhh.hh_123_band_dos}Evolution of electronic structure of HH with excess Ni: band structures (a) HH+1Ni, (b) HH+2Ni, and (c) HH+3Ni, and (d) density of states of HH with 1, 2, 3 and 4 additional Ni.}
\end{figure}

In Fig.~\ref{fig.fhhh.hh_123_band_dos}, we give the band structures and DOS of HH+$n$Ni systems for different values of $n$ for the lowest energy structures. As one can see in Fig.~\ref{fig.fhhh.hh_123_band_dos}(a,b,c), band structure of the host matrix is mainly unchanged when adding Ni to HH, except that defects split the degeneracy of host HH bands due to impurity-induced symmetry-lowering and introduce defect states into the gap. There is still some local residual symmetry which preserves the degeneracy of the valence band maximum at the $\Gamma$ point. The splitting of the HH conduction bands reduce the band gap (between HH valence band maximum and conduction bands minimum) from $\sim0.5$~eV to $\sim$0.4~eV. There are two defect states per additional Ni (Ni-eg states) located just below the HH conduction band edge. 

\begin{figure}
\begin{center}
\includegraphics[width=.5\columnwidth]{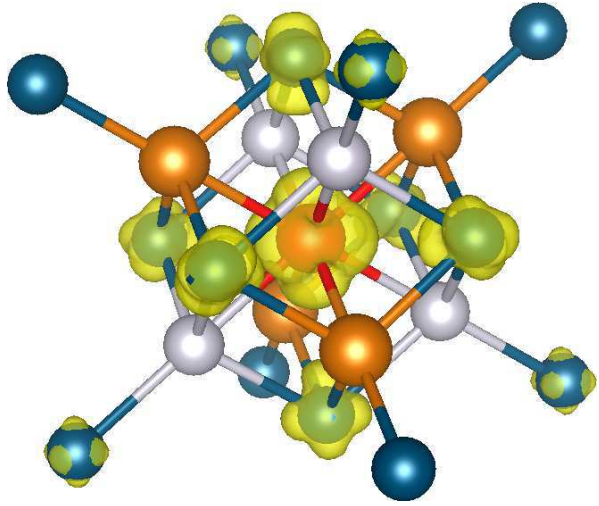}
\end{center}
\caption{\label{fig.fhhh.defect.charge}The isosurface of the charge density associated with the in-gap defect states in HH+1Ni. Host Ni are blue spheres, excess Ni is the red sphere, Zr are brown spheres, Sn are gray spheres.}
\end{figure}

When one excess Ni is added to the HH matrix, it brings 10 more electrons which occupy 5 Ni$_1$-d bands, 3 t$_{2g}$ and 2 e$_g$. All the Ni$_1$-d bands are below the Fermi level, the t$_{2g}$ bands are located around $\sim-1.5$~eV, whereas the e$_g$ bands  are located right below the HH conduction bands, giving rise to the in-gap states. The charge density associated with the in-gap defect states is given in Fig.~\ref{fig.fhhh.defect.charge}, clearly showing the Ni-e$_g$ character. The  defect bands have a bandwidth of $\sim0.14$~eV. The dispersion is small in $x$, $y$, or $z$ directions, but larger in other directions, such as $\Gamma$-M (110), $\Gamma$-R (111), which are directions of Ni$_1$-Zr(Sn) bonds. In order to understand the nature of the defect states, we have studied $3\times3\times3$ supercell with one excess Ni ($\delta=0.009$). This increases the distance between the defect and its supercell images from $\sim12$\AA{} to $\sim18$\AA. We find that the dispersion of the defect bands decreases dramatically (to almost flat bands). This result implies that the coupling between the defects is weak at extremely dilute concentration but is important for moderate and high concentrations. 

When more Ni are added to the HH matrix, more impurity bands are introduced into the gap region, and eventually they fill up the gap, converting the semiconducting HH into a metallic FH. Comparing the DOS of HH+$n$Ni (Fig.~\ref{fig.fhhh.hh_123_band_dos}d), one can clearly see the evolution of the electronic structure. With one extra Ni, DOS starts to show a peak right below the conduction bands. As the number of excess Ni increases the peak gets higher and shifts downwards in energy. These localized defect states are occupied and do not contribute directly to charge transport but can act as donor-states  injecting carriers into the conduction band. Also they can suppress $p$-type behavior by destroying the holes in the valence band. The presence of such localized states in HH-FH alloys for a small concentration of excess Ni (beyond HH limit) can dramatically affect their transport properties.

\begin{figure}
\centering
\includegraphics[width=.9\columnwidth]{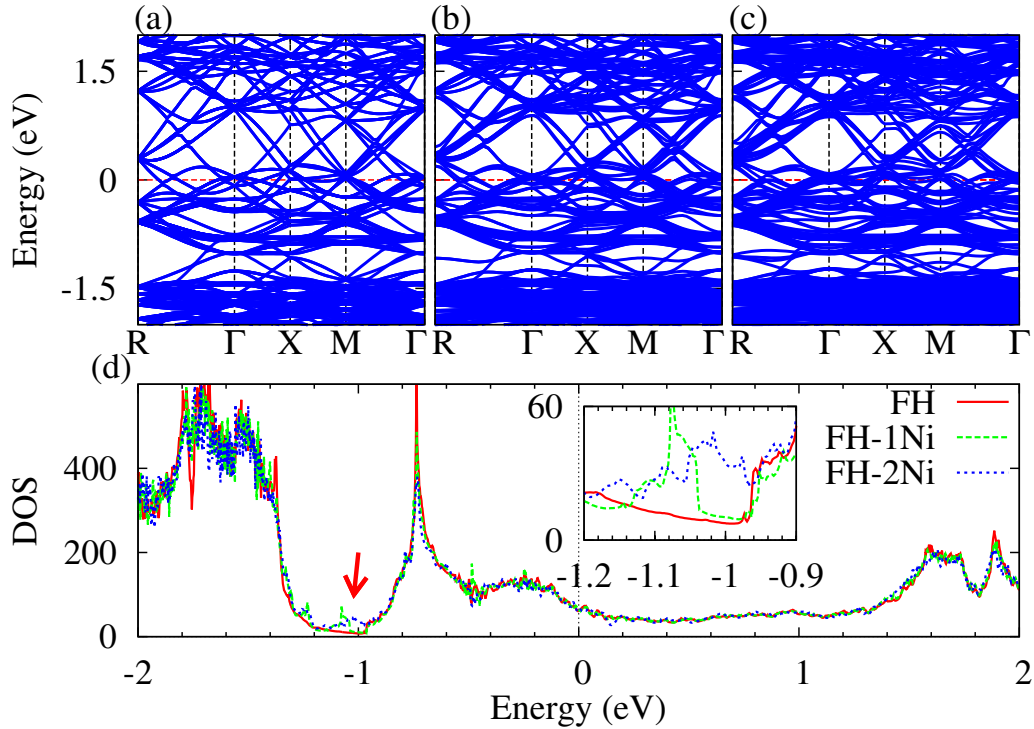}
\caption{\label{fig.fhhh.fh_012_band_dos}Evolution of electronic structure of FH with deficient Ni: band structures (a) FH, (b) FH-1Ni, and (c) FH-2Ni , and density of states of those systems. The inset show the DOS, indicated by the arrow, around -1~eV in small range [-1.2,-0.9] eV.}
\end{figure}

The effect of Ni-vacancy in FH is less subtle than that of Ni excess in HH. Fig.~\ref{fig.fhhh.fh_012_band_dos} gives the electronic structure of FH-$n$Ni. The dominant effect of Ni-vacancy on the band structure of FH is the band splitting. One can see some extra bands appearing around $\sim$-1 eV. These extra bands give rise to a small peak in the DOS at $\sim$-1 eV (Fig.~\ref{fig.fhhh.fh_012_band_dos}d inset). These effects, however, introduce small changes in the DOS (figure~\ref{fig.fhhh.fh_012_band_dos}d), particularly near the Fermi level and do not change the metallic properties of FH, at least for small concentrations of Ni-vacancy.

\section{\label{sec.hhfh.sum}Summary and Conclusion}

To summarize our work, we have discussed the energetics and electronic structures of the HH-FH mixture, ZrNi$_{1+x}$Sn, using supercell model. The focus is on the small concentration ($\lesssim0.125$) limit of defects (excess Ni in HH and Ni-vacancy in FH). We have also done some limited calculations using $3\times3\times3$.

Our calculations predict an attractive interaction between excess Ni atoms in the HH matrix, \ie the excess Ni atoms tend to come close to each other and form nano-clusters within the HH matrix. The lowest energy nanoclusters are: dimers (2 Ni), trimers (3 Ni) with equilateral triangle structure, and tetramers (4 Ni) with a distorted rhombus (DR) structure. Tetramers with tetrahedral structure which form an embryonic FH unit cell have slightly higher energy than the DR structure. In order to mimic the nanometer size FH inclusions in a HH matrix (seen by Makongo \etal\cite{makongo_simultaneous_2011}) one needs to go to larger clusters of Ni atoms, which requires much larger supercell or other non \abinitio theoretical methods.
 
Vacancies in the FH matrix, on the other hand, are found to possibly occupy either of two crystallographic sites, h(1/4,1/4,1/4)  or f(3/4,3/4,3/4), and their energetics are distinctively different from that of excess Ni in the HH matrix. If two Ni-vacancies in FH occupy the equivalent sites(either type h or f), the interaction between them is attractive at a distance larger or equal 6.32~\AA\, and repulsive at smaller distance. If two Ni-vacancies in FH occupy the sites of different types (one at h-site and another at f-site), then the interaction is repulsive, the two vacancies tend to stay away from each other.

Analysis of the change in the charge density shows that excess Ni (or Ni-vacancy) in HH (FH) causes large charge rearrangement in the host matrix. The charge rearrangement occurs mostly on the Ni-sites and is larger at Ni sites which are connected to the excess Ni (Ni-vacancy) indirectly through one channel mediated by either a Zr or a Sn atom. The charge rearrangement is modest at the Ni-sites which are connected to the defect site through two channels mediated by  atoms of the same type. FH with one Ni-vacancy has an interesting charge rearrangement at the Ni-sites connected to the vacancy through two channels, one mediated by Zr and the other by Sn. This charge rearrangement is in different directions compared to that at other Ni-sites and is most likely responsible for the difference in energetics between FH-$n$Ni and HH+$n$Ni.

While the effect of Ni-vacancies in FH on the electronic structure is modest, the excess Ni in HH changes the HH's electronic structure dramatically. The excess Ni in HH introduces new impurity states in the gap of the HH band structure, giving rise to finite density of states below the conduction bands. Due to the in-gap impurity states, HH with excess Ni can behave like a $n$-type semiconductor. 

\ack

We would like to thanks Professor Pierre P. Poudeu at University of Michigan for useful experimental perspectives and Professor Puru Jena at Virginia Commonwealth University for the discussion on Ni cluster.

This work was supported by the Center for Revolutionary Materials for Solid State Energy Conversion, an Energy Frontier Research Center funded by the U.S. Department of Energy, Office of Science, Office of Basic Energy Sciences under Award Number DE-SC0001054.

The calculations were done using computational resource provided by National Energy Research Scientific Computing Center(NERSC) and Michigan State University (MSU), Institution for Cyber Enabled Research (ICER), and High Performance Computer Center (HPCC). 
\section*{References}
\bibliography{reference,hh_fh_ref,my_publications}   % name your BibTeX data base

\begin{thebibliography}{10}

\bibitem{galanakis_introduction_2006}
I.~Galanakis, Ph~Mavropoulos, and P.~H. Dederichs.
\newblock Introduction to half-metallic heusler alloys: Electronic structure
  and magnetic properties.
\newblock {\em J. Phys. D: Appl. Phys.}, 39:765--775, March 2006.

\bibitem{galanakis_spin-polarization_2007}
I.~Galanakis and Ph~Mavropoulos.
\newblock Spin-polarization and electronic properties of half-metallic
  {Heusler} alloys calculated from first principles.
\newblock {\em J. Phys.: Condens. Matter}, 19(31):315213, August 2007.

\bibitem{katsnelson_half-metallic_2008}
M.~I. Katsnelson, V.~Yu. Irkhin, L.~Chioncel, A.~I. Lichtenstein, and R.~A.
  de~Groot.
\newblock Half-metallic ferromagnets: From band structure to many-body effects.
\newblock {\em Rev. Mod. Phys.}, 80(2):315--378, April 2008.

\bibitem{picozzi_first-principles_2008}
Silvia Picozzi.
\newblock First-principles study of ferromagnetic {Heusler} alloys: An
  overview.
\newblock In Prof Dr~Rolf Haug, editor, {\em Advances in Solid State Physics},
  number~47 in Advances in Solid State Physics, pages 129--141. Springer Berlin
  Heidelberg, January 2008.

\bibitem{graf_heusler_2011}
T.~Graf, Stuart S~P Parkin, and Claudia Felser.
\newblock Heusler compounds -- a material class with exceptional properties.
\newblock {\em {IEEE} Transactions on Magnetics}, 47(2):367--373, 2011.

\bibitem{graf_simple_2011}
Tanja Graf, Claudia Felser, and Stuart~{S.P.} Parkin.
\newblock Simple rules for the understanding of {Heusler} compounds.
\newblock {\em Progress in Solid State Chemistry}, 39(1):1--50, May 2011.

\bibitem{casper_half-heusler_2012}
F.~Casper, T.~Graf, S.~Chadov, B.~Balke, and C.~Felser.
\newblock Half-{Heusler} compounds: novel materials for energy and spintronic
  applications.
\newblock {\em Semiconductor Science and Technology}, 27(6):063001, June 2012.

\bibitem{bai_data_2013}
Zhaoqiang Bai, Lei Shen, Guchang Han, and Yuan~ping Feng.
\newblock Data storage: Review of {Heusler} compounds.
\newblock {\em {arXiv:1301.5455}}, January 2013.

\bibitem{galanakis_slater-pauling_2013}
I.~Galanakis.
\newblock {Slater}-{Pauling} behavior in half-metallic magnets.
\newblock {\em {arXiv:1302.4699}}, February 2013.

\bibitem{aliev_metal-insulator_1987}
F.~G. Aliev, {N.B.} Brandt, V.~V. Kozyr'kov, V.~V. Moshchalkov, R.~V.
  Skolozdra, Yu.~V. Stadnyk, and V.~K. Pecharski.
\newblock Metal-insulator transition of {RNiSn} ({R=Zr,Hf,Ti)} intermetallic
  vacancy system.
\newblock {\em Pis'ma Zhr. Eksp. Teor. Fiz.}, 45(11):535--537, 1987.

\bibitem{aliev_narrow_1990}
F.~G. Aliev, V.~V. Kozyrkov, V.~V. Moshchalkov, R.~V. Scolozdra, and
  K.~Durczewski.
\newblock Narrow band in the intermetallic compounds {MNiSn} ({M=Ti}, {Zr,
  Hf}).
\newblock {\em Zeitschrift f\"ur Physik B Condensed Matter}, 80(3):353--357,
  October 1990.

\bibitem{aliev_gap_1991}
{F.G.} Aliev.
\newblock Gap at {Fermi} level in some new d- and f-electron intermetallic
  compounds.
\newblock {\em Physica B: Condens. Matter}, 171(1–4):199--205, May 1991.

\bibitem{aliev_anomalous_1993}
F.~G. Aliev, V.~V. Pryadun, R.~Villar, S.~Vieira, J.~M. Calleja, N.~Mestres,
  and R.~V. Scolozdra.
\newblock Anomalous lattice properties of {ZrNiSn} caused by electron
  localization.
\newblock {\em Int. J. Mod. Phys. B}, 07(01n03):383--386, January 1993.

\bibitem{mestres_electron_1994}
N.~Mestres, {J.M.} Calleja, {F.G.} Aliev, and {A.I.} Belogorokhov.
\newblock Electron localization in the disordered conductors {TiNiSn} and
  {HfNiSn} observed by raman and infrared spectroscopies.
\newblock {\em Solid State Commun.}, 91(10):779--784, September 1994.

\bibitem{slebarski_electronic_1998}
A.~Slebarski, A.~Jezierski, S.~L\"utkehoff, and M.~Neumann.
\newblock Electronic structure of {X}$_{2}${ZrSn-} and {XZrSn-type} {Heusler}
  alloys with {X=Co} or {Ni}.
\newblock {\em Phys. Rev. B}, 57(11):6408--6412, March 1998.

\bibitem{hohl_efficient_1999}
Heinrich Hohl, Art~P. Ramirez, Claudia Goldmann, Gabriele Ernst, Bernd
  W\"olfing, and Ernst Bucher.
\newblock Efficient dopants for {ZrNiSn-based} thermoelectric materials.
\newblock {\em J. Phys.: Condens. Matter}, 11(7):1697, 1999.

\bibitem{larson_structural_2000}
Paul Larson, S.~D. Mahanti, and M.~G. Kanatzidis.
\newblock Structural stability of {Ni}-containing half-{Heusler} compounds.
\newblock {\em Phys. Rev. B}, 62(19):12754, 2000.

\bibitem{shen_thermoelectric_2001}
Qiang Shen, Lianmeng Zhang, Lidong Chen, Takashi Goto, and Toshio Hirai.
\newblock Thermoelectric properties of {ZrNiSn-based} half-{Heusler} compounds
  by solid state reaction method.
\newblock {\em J. Mater. Sci. Lett.}, 20(24):2197–2199, 2001.

\bibitem{germond_thermoelectric_2010}
Jeffrey~D Germond, Paul~J Schilling, Nathan~J. Takas, and Pierre F.~P. Poudeu.
\newblock Thermoelectric performance of nanostructured {ZrNiSn} compounds
  synthesized by mechanical alloying.
\newblock {\em {MRS} Proc.}, 1267, 2010.

\bibitem{kimura_vacancy_2010}
Yoshisato Kimura, Toshiyasu Tanoguchi, and Takuji Kita.
\newblock Vacancy site occupation by {Co} and {Ir} in half-{Heusler} {ZrNiSn}
  and conversion of the thermoelectric properties from n-type to p-type.
\newblock {\em Acta Materialia}, 58(13):4354--4361, August 2010.

\bibitem{qiu_effect_2010}
Pengfei Qiu, Jiong Yang, Xiangyang Huang, Xihong Chen, and Lidong Chen.
\newblock Effect of antisite defects on band structure and thermoelectric
  performance of {ZrNiSn} half-{Heusler} alloys.
\newblock {\em Appl. Phys. Lett.}, 96(15):152105--152105--3, April 2010.

\bibitem{kimura_effect_2011}
Yoshisato Kimura, Toshiyasu Tanoguchi, Yasuhiro Sakai, Yaw-Wang Chai, and
  Yoshinao Mishima.
\newblock Effect of vacancy-site occupation in half-{Heusler} compound {ZrNiSn}
  on phase stability and thermoelectric properties.
\newblock {\em {MRS} Proc.}, 1295, March 2011.

\bibitem{lee_validity_2012}
Mal-Soon Lee and S.~D. Mahanti.
\newblock Validity of the rigid band approximation in the study of the
  thermopower of narrow band gap semiconductors.
\newblock {\em Phys. Rev. B}, 85(16):165149, April 2012.

\bibitem{chen_effect_2013}
Shuo Chen, Kevin~C. Lukas, Weishu Liu, Cyril~P. Opeil, Gang Chen, and Zhifeng
  Ren.
\newblock Effect of {Hf} concentration on thermoelectric properties of
  nanostructured n-type half-{Heusler} materials
  {Hf$_x$Zr$_{1–x}$NiSn$_{0.99}$Sb$_{0.01}$}.
\newblock {\em Adv. Energy Mater.}, 2013.

\bibitem{romaka_peculiarities_2013}
{V.V.} Romaka, P.~Rogl, L.~Romaka, Yu. Stadnyk, A.~Grytsiv, O.~Lakh, and
  V.~Krayovskii.
\newblock Peculiarities of structural disorder in {Zr}- and {Hf}-containing
  {Heusler} and half-{Heusler} stannides.
\newblock {\em Intermetallics}, 35:45--52, April 2013.

\bibitem{zou_electronic_2013}
D.~F. Zou, S.~H. Xie, Y.~Y. Liu, J.~G. Lin, and J.~Y. Li.
\newblock Electronic structure and thermoelectric properties of half-{Heusler}
  {Zr$_{0.5}$Hf$_{0.5}$NiSn} by first-principles calculations.
\newblock {\em J. Appl. Phys.}, 113(19):193705--7, May 2013.

\bibitem{sakurada_effect_2005}
S.~Sakurada and N.~Shutoh.
\newblock Effect of {Ti} substitution on the thermoelectric properties of
  ({Zr,Hf)NiSn} half-{Heusler} compounds.
\newblock {\em Appl. Phys. Lett.}, 86(8):082105, February 2005.

\bibitem{chaput_electronic_2006}
L.~Chaput, J.~Tobola, P.~P\'echeur, and H.~Scherrer.
\newblock Electronic structure and thermopower of {Ni(Ti$_{0.5}$Hf$_{0.5}$)Sn}
  and related half-{Heusler} phases.
\newblock {\em Phys. Rev. B}, 73(4):045121, January 2006.

\bibitem{kim_high_2007}
Sung-Wng Kim, Yoshisato Kimura, and Yoshinao Mishima.
\newblock High temperature thermoelectric properties of {TiNiSn-based}
  half-{Heusler} compounds.
\newblock {\em Intermetallics}, 15(3):349--356, March 2007.

\bibitem{wang_thermoelectric_2009}
L.~L. Wang, L.~Miao, Z.~Y. Wang, W.~Wei, R.~Xiong, H.~J. Liu, J.~Shi, and X.~F.
  Tang.
\newblock Thermoelectric performance of half-{Heusler} compounds {TiNiSn} and
  {TiCoSb}.
\newblock {\em J. Appl. Phys.}, 105(1):013709--013709--5, January 2009.

\bibitem{yu_high-performance_2009}
Cui Yu, Tie-Jun Zhu, Rui-Zhi Shi, Yun Zhang, Xin-Bing Zhao, and Jian He.
\newblock High-performance half-{Heusler} thermoelectric materials
  {Hf$_{1−x}$Zr$_x$NiSn$_{1−y}$Sb$_y$} prepared by levitation melting and
  spark plasma sintering.
\newblock {\em Acta Materialia}, 57(9):2757--2764, May 2009.

\bibitem{hazama_improvement_2011}
Hirofumi Hazama, Masato Matsubara, Ryoji Asahi, and Tsunehiro Takeuchi.
\newblock Improvement of thermoelectric properties for half-{Heusler} {TiNiSn}
  by interstitial {Ni} defects.
\newblock {\em Journal of Applied Physics}, 110(6):063710--063716, September
  2011.

\bibitem{makongo_simultaneous_2011}
Julien P.~A. Makongo, Dinesh~K. Misra, Xiaoyuan Zhou, Aditya Pant, Michael~R.
  Shabetai, Xianli Su, Ctirad Uher, Kevin~L. Stokes, and Pierre F.~P. Poudeu.
\newblock Simultaneous large enhancements in thermopower and electrical
  conductivity of bulk nanostructured half-{Heusler} alloys.
\newblock {\em J. Am. Chem. Soc.}, 133(46):18843--18852, November 2011.

\bibitem{makongo_thermal_2011}
Julien~{P.A.} Makongo, Dinesh~K. Misra, James~R. Salvador, Nathan~J. Takas,
  Guoyu Wang, Michael~R. Shabetai, Aditya Pant, Pravin Paudel, Ctirad Uher,
  Kevin~L. Stokes, and Pierre~{F.P.} Poudeu.
\newblock Thermal and electronic charge transport in bulk nanostructured
  {Zr$_{0.25}$Hf$_{0.75}$NiSn} composites with full-{Heusler} inclusions.
\newblock {\em Journal of Solid State Chemistry}, 184(11):2948--2960, November
  2011.

\bibitem{poon_half-heusler_2011}
{S.J.} Poon, D.~Wu, S.~Zhu, W.~Xie, {T.M.} Tritt, P.~Thomas, and
  R.~Venkatasubramanian.
\newblock Half-{Heusler} phases and nanocomposites as emerging high-{ZT}
  thermoelectric materials.
\newblock {\em Journal of Materials Research}, 26(22):2795--2802, 2011.

\bibitem{douglas_enhanced_2012}
Jason~E. Douglas, Christina~S. Birkel, Mao-Sheng Miao, Chris~J. Torbet,
  Galen~D. Stucky, Tresa~M. Pollock, and Ram Seshadri.
\newblock Enhanced thermoelectric properties of bulk {TiNiSn} via formation of
  a {TiNi$_2$Sn} second phase.
\newblock {\em App. Phys. Lett.}, 101(18):183902--183904, nov 2012.

\bibitem{wang_chai_nanosized_2012}
Yaw~Wang Chai and Yoshisato Kimura.
\newblock Nanosized precipitates in half-{Heusler} {TiNiSn} alloy.
\newblock {\em Appl. Phys. Lett.}, 100(3):033114, January 2012.

\bibitem{joshi_enhancement_2013}
Giri Joshi, Tulashi Dahal, Shuo Chen, Hengzhi Wang, Junichiro Shiomi, Gang
  Chen, and Zhifeng Ren.
\newblock Enhancement of thermoelectric figure-of-merit at low temperatures by
  titanium substitution for hafnium in n-type half-{Heuslers}
  {Hf$_{0.75−x}$Ti$_x$Zr$_{0.25}$NiSn$_{0.99}$Sb$_{0.01}$}.
\newblock {\em Nano Energy}, 2(1):82--87, January 2013.

\bibitem{appel_effects_2013}
O.~Appel, M.~Schwall, D.~Mogilyansky, M.~K\"ohne, B.~Balke, and Y.~Gelbstein.
\newblock Effects of microstructural evolution on the thermoelectric properties
  of spark-plasma-sintered {Ti$_{0.3}$Zr$_{0.35}$Hf$_{0.35}$NiSn}
  half-{Heusler} compound.
\newblock {\em Journal of Electronic Materials}, 42(7):1340--1345, July 2013.

\bibitem{birkel_improving_2013}
Christina~S. Birkel, Jason~E. Douglas, Bethany~R. Lettiere, Gareth Seward,
  Nisha Verma, Yichi Zhang, Tresa~M. Pollock, Ram Seshadri, and Galen~D.
  Stucky.
\newblock Improving the thermoelectric properties of half-{Heusler} {TiNiSn}
  through inclusion of a second full-{Heusler} phase: microwave preparation and
  spark plasma sintering of {TiNi$_{1+x}$Sn}.
\newblock {\em Physical Chemistry Chemical Physics}, 15(18):6990, 2013.

\bibitem{kirievsky_phase_2013}
K.~Kirievsky, Y.~Gelbstein, and D.~Fuks.
\newblock Phase separation and antisite defects in the thermoelectric {TiNiSn}
  half-{Heusler} alloys.
\newblock {\em Journal of Solid State Chemistry}, 203:247--254, July 2013.

\bibitem{yan_thermoelectric_2013}
Xiao Yan, Weishu Liu, Shuo Chen, Hui Wang, Qian Zhang, Gang Chen, and Zhifeng
  Ren.
\newblock Thermoelectric property study of nanostructured p-type
  half-{Heuslers} {(Hf, Zr, Ti)CoSb$_{0.8}$Sn$_{0.2}$}.
\newblock {\em Adv. Energy Mater.}, 2013.

\bibitem{downie_enhanced_2013}
R.~A. Downie, D.~A. {MacLaren}, R.~I. Smith, and J.~W.~G. Bos.
\newblock Enhanced thermoelectric performance in {TiNiSn-based}
  half-{Heuslers}.
\newblock {\em Chem. Commun.}, 49(39):4184--4186, April 2013.

\bibitem{faleev_en_filter_2008}
Sergey~V. Faleev and Fran\c{c}ois L\'eonard.
\newblock Theory of enhancement of thermoelectric properties of materials with
  nanoinclusions.
\newblock {\em Phys. Rev. B}, 77(21):214304, June 2008.

\bibitem{biswas_strained_2011}
Kanishka Biswas, Jiaqing He, Qichun Zhang, Guoyu Wang, Ctirad Uher, Vinayak~P.
  Dravid, and Mercouri~G. Kanatzidis.
\newblock Strained endotaxial nanostructures with high thermoelectric figure of
  merit.
\newblock {\em Nat. Chem.}, 3(2):160--166, February 2011.

\bibitem{stadnyk_isothermal_2005}
Yu. Stadnyk, L.~Romaka, A.~Horyn, A.~Tkachuk, Yu. Gorelenko, and P.~Rogl.
\newblock Isothermal sections of the {Ti–Co–Sn} and {Ti–Co–Sb} systems.
\newblock {\em Journal of Alloys and Compounds}, 387(1–2):251--255, January
  2005.

\bibitem{akai_kkr_1989}
H~Akai.
\newblock Fast korringa-kohn-rostoker coherent potential approximation and its
  application to fcc {Ni-Fe} systems.
\newblock {\em Journal of Physics: Condensed Matter}, 1(43):8045, 1989.

\bibitem{defects}
Risto~M. Nieminen.
\newblock {\em Topics in Applied Physics: Theory of defects in semiconductors},
  volume 104, pages 36--40.
\newblock Springer, 2006.

\bibitem{do_fe2val.2011}
Dat Do, Mal-Soon Lee, and S.~D. Mahanti.
\newblock Effect of onsite coulomb repulsion on thermoelectric properties of
  full-heusler compounds with pseudogaps.
\newblock {\em Phys. Rev. B}, 84:125104, Sep 2011.

\bibitem{do_cusbse.2012}
Dat Do, Vidvuds Ozolins, S~D Mahanti, Mal-Soon Lee, Yongsheng Zhang, and
  C~Wolverton.
\newblock Physics of bandgap formation in {Cu–Sb–Se} based novel
  thermoelectrics: the role of {Sb} valency and {Cu} d levels.
\newblock {\em J. Phys.: Condens. Matter}, 24(41):415502, 2012.

\bibitem{ogut_bandgap_stability_1995}
Serdar \"O\ifmmode~\breve{g}\else \u{g}\fi{}\"ut and Karin~M. Rabe.
\newblock Band gap and stability in the ternary intermetallic compounds
  {NiSn\textit{M}(\textit{M}=Ti,Zr,Hf)}: {A} first-principles study.
\newblock {\em Phys. Rev. B}, 51:10443--10453, Apr 1995.

\bibitem{mbj}
Fabien Tran and Peter Blaha.
\newblock Accurate band gaps of semiconductors and insulators with a semilocal
  exchange-correlation potential.
\newblock {\em Phys. Rev. Lett.}, 102(22):226401, Jun 2009.

\bibitem{hse06:1}
Jochen Heyd, Gustavo~E. Scuseria, and Matthias Ernzerhof.
\newblock Hybrid functionals based on a screened coulomb potential.
\newblock {\em The Journal of Chemical Physics}, 118:8207, 2003.

\bibitem{hse06:2}
Jochen Heyd and Gustavo~E. Scuseria.
\newblock Efficient hybrid density functional calculations in solids:
  Assessment of the {Heyd--Scuseria-–Ernzerhof} screened coulomb hybrid
  functional.
\newblock {\em The Journal of Chemical Physics}, 121:1187, 2004.

\bibitem{hse06:3}
Jochen Heyd, Gustavo~E. Scuseria, and Matthias Ernzerhof.
\newblock Erratum: {“Hybrid} functionals based on a screened coulomb
  potential” {[J.} chem. phys. 118, 8207 (2003)].
\newblock {\em The Journal of Chemical Physics}, 124:219906, 2006.

\bibitem{pbe}
John~P. Perdew, Kieron Burke, and Matthias Ernzerhof.
\newblock Generalized gradient approximation made simple.
\newblock {\em Phys. Rev. Lett.}, 77(18):3865--3868, Oct 1996.

\bibitem{bloch94}
P.~E. Bl\"ochl.
\newblock {\em Phys. Rev. B}, 50:17953--17979, 1994.

\bibitem{kresse99}
G.~Kresse and D.~Joubert.
\newblock From ultrasoft pseudopotentials to the projector augmented-wave
  method.
\newblock {\em Phys. Rev. B}, 59(3):1758--1775, Jan 1999.

\bibitem{vasp1}
G.~Kresse and J.~Hafner.
\newblock Ab initio molecular dynamics for liquid metals.
\newblock {\em Phys. Rev. B}, 47(1):558--561, Jan 1993.

\bibitem{vasp2}
G.~Kresse and J.~Furthm\"uller.
\newblock Efficiency of ab-initio total energy calculations for metals and
  semiconductors using a plane-wave basis set.
\newblock {\em Computational Materials Science}, 6(1):15 -- 50, 1996.

\bibitem{vasp3}
G.~Kresse and J.~Furthm\"uller.
\newblock Efficient iterative schemes for ab initio total-energy calculations
  using a plane-wave basis set.
\newblock {\em Phys. Rev. B}, 54(16):11169--11186, Oct 1996.

\bibitem{monkhorst76}
Hendrik~J. Monkhorst and James~D. Pack.
\newblock Special points for brillouin-zone integrations.
\newblock {\em Phys. Rev. B}, 13(12):5188--5192, Jun 1976.

\bibitem{zhang_defect_cuinse_1998}
S.~B. Zhang, Su-Huai Wei, Alex Zunger, and H.~Katayama-Yoshida.
\newblock Defect physics of the {CuInSe}$_{2}$ chalcopyrite semiconductor.
\newblock {\em Phys. Rev. B}, 57:9642--9656, Apr 1998.

\bibitem{menon_tb_md.1994}
Madhu Menon, John Connolly, Nectarios Lathiotakis, and Antonis Andriotis.
\newblock Tight-binding molecular-dynamics study of transition-metal clusters.
\newblock {\em Phys. Rev. B}, 50:8903--8906, Sep 1994.

\bibitem{Reuse_Nicluster.1995}
F.A. Reuse and S.N. Khanna.
\newblock Geometry, electronic structure, and magnetism of small nin (n =
  2–6, 8, 13) clusters.
\newblock {\em Chemical Physics Letters}, 234(1–3):77 -- 81, 1995.

\bibitem{Jena_energetics.1997}
Saroj~K. Nayak, S.~N. Khanna, B.~K. Rao, and P.~Jena.
\newblock Physics of nickel clusters:  energetics and equilibrium geometries.
\newblock {\em The Journal of Physical Chemistry A}, 101(6):1072--1080, 1997.

\bibitem{valeri_Nien.2004}
Valeri~G. Grigoryan and Michael Springborg.
\newblock Structural and energetic properties of nickel clusters:$2\le n\le
  150$.
\newblock {\em Phys. Rev. B}, 70:205415, Nov 2004.

\bibitem{futscheck_abinitio.2006}
T~Futschek, J~Hafner, and M~Marsman.
\newblock Stable structural and magnetic isomers of small transition-metal
  clusters from the ni group: an ab initio density-functional study.
\newblock {\em Journal of Physics: Condensed Matter}, 18(42):9703, 2006.

\bibitem{lu.2007}
Q.~L. Lu, Q.~Q. Luo, L.~L. Chen, and J.~G. Wan.
\newblock Structural and magnetic properties of ni n (n = 2–21) clusters.
\newblock {\em The European Physical Journal D}, 61(2):389--396, 2011.

\bibitem{Deshpande.2005}
Mrinalini Deshpande, D.~G. Kanhere, and Ravindra Pandey.
\newblock Structures, energetics, and magnetic properties of ni$_n$b clusters
  with $n=1-8,12$.
\newblock {\em Phys. Rev. A}, 71:063202, Jun 2005.

\bibitem{petkov.2006}
Petko St.~Petkov, Georgi~N. Vayssilov, Sven Kruger, and Notker Rosch.
\newblock Structure{,} stability{,} electronic and magnetic properties of ni4
  clusters containing impurity atoms.
\newblock {\em Phys. Chem. Chem. Phys.}, 8:1282--1291, 2006.

\bibitem{deshpande.2007}
M.~D. Deshpande, Swapna Roy, and D.~G. Kanhere.
\newblock Equilibrium geometries, electronic structure, and magnetic properties
  of ni$_n$sn clusters ($n=1-12$).
\newblock {\em Phys. Rev. B}, 76:195423, Nov 2007.

\bibitem{chikaoui.2011}
A.~Chikhaoui, K.~Haddab, S.~Bouarab, and A.~Vega.
\newblock Density functional study of the structures and electronic properties
  of nitrogen-doped nin clusters, n = 1–10.
\newblock {\em The Journal of Physical Chemistry A}, 115(48):13997--14005,
  2011.

\bibitem{larson_electronic_1999}
P.~Larson, S.~D. Mahanti, Sandrine Sportouch, and M.~G. Kanatzidis.
\newblock Electronic structure of rare-earth nickel pnictides: Narrow-gap
  thermoelectric materials.
\newblock {\em Phys. Rev. B}, 59(24):15660, 1999.

\bibitem{winterlik.superonductor.2008}
J\"urgen Winterlik, Gerhard~H. Fecher, Claudia Felser, Martin Jourdan, Kai
  Grube, Fr\'ed\'eric Hardy, Hilbert von L\"ohneysen, K.~L. Holman, and R.~J.
  Cava.
\newblock Ni-based superconductor: Heusler compound {ZrNi$_{2}$Ga}.
\newblock {\em Phys. Rev. B}, 78:184506, Nov 2008.

\bibitem{boff_specificheat.1996}
M.~A.~S. Boff, G.~L.~F. Fraga, D.~E. Brand\~ao, A.~A. Gomes, and T.~A. Grandi.
\newblock Specific heat of {Ni$_2$TSn (T = Ti, Zr, Hf)} heusler compounds.
\newblock {\em physica status solidi (a)}, 154(2):549--552, 1996.

\end{thebibliography}

\end{document}